# Electrodeposited WS$_2$ Monolayers on Fabricated Graphene Electrodes

Y. J. Noori[1,5,*], S. Thomas[2], S. Ramadan[3], V. K. Greenacre[2], N. M. Abdelazim[1], Y. Han[4], J. Zhang[1], R. Beanland[4], A. L. Hector[2], N. Klein[3], G. Reid[2], P. N. Bartlett[2] and C. H. de Groot[1,*]

[1]School of Electronics and Computer Science, University of Southampton, Southampton, SO17 1BJ, UK
[2]School of Chemistry, University of Southampton, Southampton, SO17 1BJ, UK
[3]Department of Materials, Imperial College, London, SW7 2AZ, UK
[4]Department of Physics, University of Warwick, Coventry, CV4 7AL, UK
[5]Optoelectronic Research Centre, University of Southampton, Southampton, SO17 1BJ, UK

* Email: y.j.noori@southampton.ac.uk; chdg@soton.ac.uk

## Abstract

The development of scalable techniques to make 2D material heterostructures is a major obstacle that needs to be overcome before these materials can be implemented in device technologies industrially. Electrodeposition is an industrially compatible deposition technique that offers unique advantages in scaling 2D heterostructures. In this work, we demonstrate the electrodeposition of atomic layers of WS$_2$ over graphene electrodes using a single source precursor. Using conventional microfabrication techniques, graphene was patterned to create micro-electrodes where WS$_2$ was site-selectively deposited to form 2D heterostructures. We used various characterisation techniques, including atomic force microscopy, transmission electron microscopy, Raman spectroscopy and x-ray photoelectron spectroscopy to show that our electrodeposited WS$_2$ layers are highly uniform and can be grown over graphene at a controllable deposition rate. This technique to selectively deposit TMDCs over microfabricated graphene electrodes paves the way towards wafer-scale production of 2D material heterostructures for nanodevice applications.

**Keywords:** Electrodeposition, 2D materials, non-aqueous, transition metal dichalcogenides, TMDC, tungsten disulfide, WS$_2$, graphene, heterostructure

## Introduction

The exceptional optoelectronic properties of 2D transition metal dichalcogenides (TMDCs) have made these materials a hot research topic since the early 2010s. Among the TMDC family, tungsten disulfide (WS$_2$) stands out for its high quantum yield which makes it exhibit strong photoluminescence, as well as its high electrical performance and ambipolar field effect behaviour.[1–6] Emerging technological applications of tungsten disulfide include bright light emitting devices, high responsivity photodetectors, sensitive gas sensors and optical identification labels.[7–9] Heterostructures comprising of several 2D materials can harness the unique properties of their individual material constituents to make new device technologies that have the potential to surpass the performances of those commercialised and reported in the literature. For example, several papers have demonstrated photodetection by combining the unique semiconductor properties of WS$_2$ with that of graphene, offering photodetectors with broadband detection capability and high photoresponsivity.[10–13]

The challenges associated with scaling the production of 2D material heterostructures present a major obstacle that hinders their adoption in industry. There are several scalable techniques that have been used to deposit WS$_2$, such as chemical vapour deposition (CVD), atomic layer deposition, plasma sputtering, pulsed laser deposition, thermal decomposition, liquid exfoliation and electrodeposition.[4, 13–18] Among these CVD stands out because it can controllably deposit monolayer films of very high crystallinity.[19] However, 2D heterostructures grown via CVD are usually limited in area and are substrate specific.[20] CVD is also not typically a site selective deposition technique and its uses in growing TMDCs usually require elevating the host substrates to temperatures that exceed 800 ºC, which damage 2D materials such as graphene existing on the substrate.[21] These issues also limit the practicality of CVD in fabricating large numbers of 2D material heterostructure devices on each chip. Plasma based deposition techniques are also known to cause ion bombardment of the substrate, which causes crystal defects on other materials such as graphene.[22, 23]

In comparison, electrodeposition is a room temperature technique that lends itself to depositing materials selectively over conductive electrodes. It is used routinely in the semiconductor industry such as in depositing magnetic films for hard drive read/write heads, or in creating Cu chip interconnects, using the damascene process.[24, 25] We have previously used electrodeposition to grow few-layer MoS$_2$ films.[26, 27] A few works have demonstrated the electrodeposition of bulk WS$_2$ thin films over traditional electrodes such as platinum and indium tin oxide using aqueous and acetonitrile solutions.[28–30]. [NH$_4$]$_2$[WS$_4$] was commonly used as the precursor in these studies. In our recent work, we have shown that WS$_2$ thin films can be electrodeposited using dichloromethane (CH$_2$Cl$_2$), over TiN electrodes. Electrodepositing from CH$_2$Cl$_2$ and avoiding the need for addition of a proton source (that is necessary to remove the excess S when using [WS$_4$]$^{2-}$ as the precursor) increases the faradaic efficiency and extends the accessible potential window for depositing a variety of technologically important materials that may not otherwise be possible using aqueous or acetonitrile electrolytes.[31, 32] CH$_2$Cl$_2$ is also relatively inert, less viscous than ionic liquids and is only weakly coordinating. We have performed our deposition using a single source precursor based on [NEt$_4$]$_2$[WS$_2$Cl$_4$]. Compared to [NH$_4$]$_2$[WS$_4$], the advantage of this precursor is that it can be dissolved readily in CH$_2$Cl$_2$ and contains the correct W:S ratio for WS$_2$, so there is no excess S to be removed.

Challenges associated with graphene production at wafer scales have already been solved using CVD and the wet transfer technique. Several groups have demonstrated wafer-size graphene and several spin-out companies have recently emerged to exploit the commercial opportunities from this advancement, such as Graphenea Inc and Grolltex Inc.[33–35] By fabricating graphene micro-electrodes at wafer scales, one can use the scalability advantages offered by electrodeposition to site-

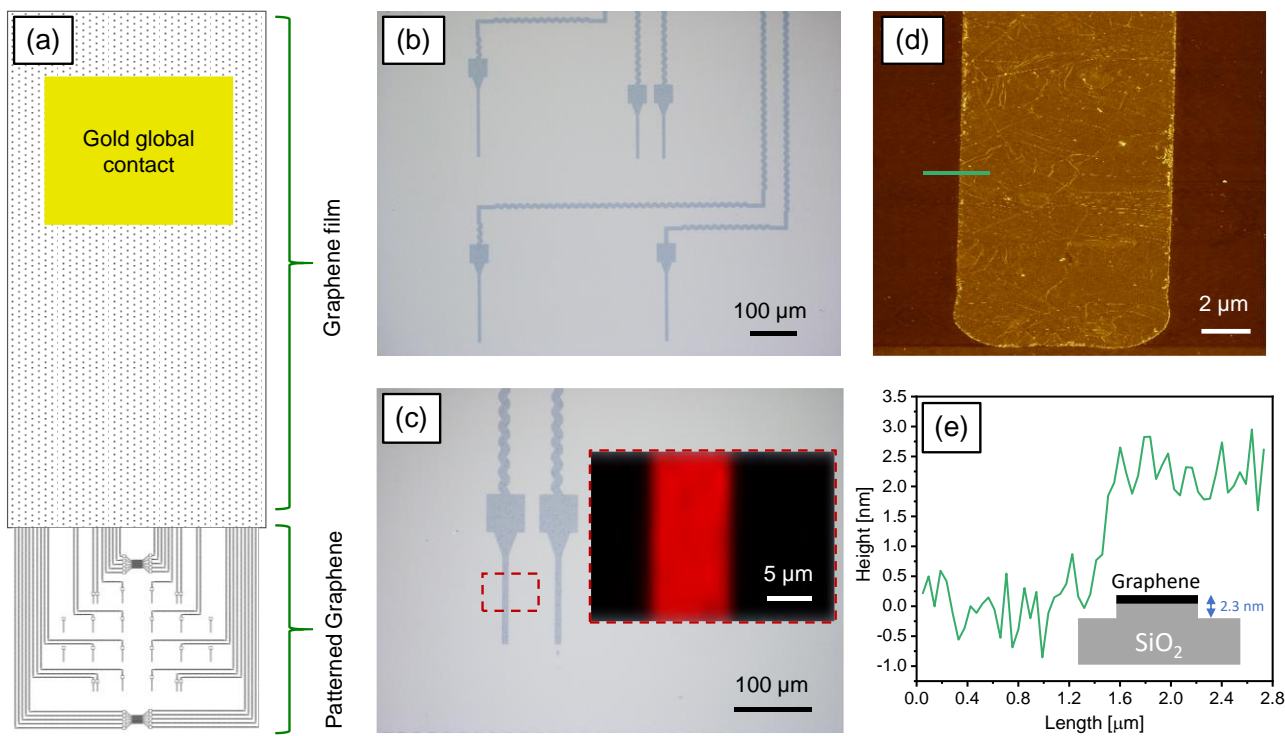

Figure 1 (a) Schematic diagram of the photolithography mask that was used in patterning the graphene electrodes. (b) Optical microscope image showing the graphene electrodes after fabrication. (c) Optical microscope image showing the graphene electrodes at higher magnification with a Raman map (inset) of the bare graphene. (d) AFM image of a graphene electrode after fabrication and a step height profile (e) taken at the edge of the electrode.

selectively deposit a variety of TMDCs to form 2D material heterostructures for device applications.

There are two aspects to the novelty of this work. Firstly, it represents the first electrochemical deposition of atomically thin $WS_2$ films over graphene electrodes. Most importantly, the deposited films presented here show far superior material uniformity and continuity compared to other electrodeposited $WS_2$ films from previous reports.[28–30] Secondly, we have shown that we can make microscale $WS_2$/graphene heterostructures by patterning the graphene electrodes using conventional microfabrication techniques. This represents a major step towards scaling the production of 2D materials and their heterostructures into wafer sizes via electroplating. Throughout the work we present results from characterising the electrodeposited $WS_2$ using various techniques including atomic force microscopy (AFM), transmission electron microscopy (TEM), Raman spectroscopy and X-ray photoelectron spectroscopy.

## Electrode Microfabrication

The graphene films used in this work were prepared by CVD on Cu foils and transferred to the Si/SiO$_2$ substrates through the wet transfer method. Details of this process has been reported previously.[36, 37] A chrome/gold pad was deposited over part of the graphene film to reduce the contact resistance between the graphene and the contact to the potentiostat.

Monolayer graphene micro-electrodes were fabricated using conventional microfabrication techniques. A photoresist was spin-coated on the graphene over a Si/SiO$_2$ substrate. The pattern was transferred to the graphene film via UV lithography and reactive ion etching (RIE) using the photoresist as an etching mask. Graphene RIE etching was performed in an O$_2$ plasma for 90 s. While shorter etching periods can be used to pattern the monolayer graphene electrodes, over etching was done to ensure that all exposed graphene or residues of photoresists are removed completely. The photoresist was then removed to expose the protected graphene micro-electrodes. Figure 1 (a) shows the layout of the photolithography mask used in this work. The resultant graphene electrodes are presented in Figure 1 (b) and (c). The smallest feature of the fabricated electrodes was approximately 9 µm. A Raman map of the graphene micro-electrodes was recorded by scanning through part of the electrodes and plotting the intensity of the graphene 2D peak (2691 cm$^{-1}$) as shown in the inset of Figure 1 (c). A typical Raman spectrum recorded from graphene electrodes is shown in Figure S1 in the supporting information. Based on the 2D/G Raman intensity ratio, and the optical microscope images, the graphene micro-electrodes should be composed of a monolayer.[38–40] The Raman spectrum shows that the quality of the graphene remains high after the fabrication process. AFM images of the graphene micro-electrodes are presented in Figure 1 (d-e) and Figure S2. AFM height analysis of the graphene electrodes shows that the electrodes are on average ~2.3 nm thick. While this thickness might appear higher than that expected for monolayer graphene, this is due to the relatively long (90s) RIE etching used, which also resulted in small amount of SiO$_2$ being physically etched in the oxygen plasma at roughly 1-2 nm/min, creating a SiO$_2$ step lower than graphene, see inset in Figure 1 (e).

## Electrodeposition of WS$_2$

Figure 2 (a) illustrates the concept of this work showing WS$_2$ 2D material over fabricated graphene strips. The graphene is connected to a large Au pad through which an electrochemical potential is applied. All the electrodeposition experiments, including the electrolyte preparation were carried out inside a glovebox equipped with a nitrogen recirculation system to maintain its O$_2$ and H$_2$O levels below 10 ppm. The electrodeposition processes were performed using a three-

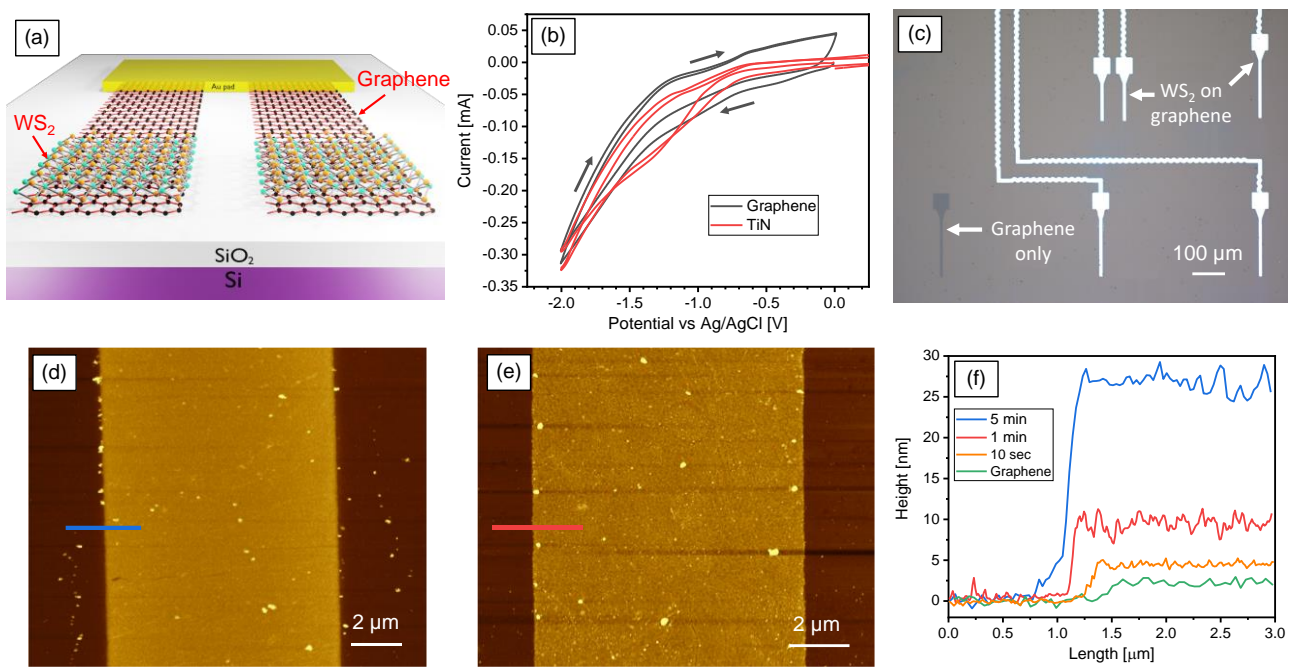

Figure 2 (a) A schematic illustration of the concept of this work showing $WS_2$ 2D materials deposited on patterned graphene stripes that are biased through an Au contact pad. (b) Cyclic voltammetry scans of the $[NEt_4]_2[WS_2Cl_4]$ precursor in DCM taken using graphene and TiN electrodes. (c) An optical microscope image showing the contrast difference between a patterned $WS_2$/graphene heterostructure and bare graphene film. AFM topography images of $WS_2$ deposits on graphene for 5 min (d) and 1 min (e). (f) Line profiles taken at the edges of the deposited films showing the total measured step heights.

electrode electrochemical cell that comprises a Pt/Ir (90:10 elemental composition) disc as the counter electrode and a reference electrode made of an Ag/AgCl wire placed within a glass frit containing 0.1 M $[N^nBu_4]Cl$ in $CH_2Cl_2$. The single source precursor, $[NEt_4]_2[WS_2Cl_4]$, was dissolved in $CH_2Cl_2$ containing 0.1 M containing $[N^nBu_4]Cl$. Details of the synthesis of this precursor has been reported previously.[41] Prior to the experiment, the $CH_2Cl_2$ was dried and degassed by distillation from $CaH_2$ to minimize its water content. The $H_2O$ content was measured by Karl-Fischer titration to be ca. 18 ppm.

The electrochemical behaviour of the precursor in $CH_2Cl_2$ was studied by performing cyclic voltammetry as shown in Figure 2 (b). The graphene and TiN electrodes chosen for the CV experiments were circularly shaped with a 4 mm diameter to ensure that a good electrical signal is obtained for the CV comparison. The potential was swept twice from 0 V to -2 V and back to 0 V as indicated by the arrows (the sweep was extended further to 1 V for the TiN electrode). The observed current range was found to be similar for both electrodes indicating that the graphene has very similar iR drop to a common conductive material like TiN. Both scans were also found to be similar. The current in both cases increases linearly as the potential was swept towards more negative potentials before a change in the gradient occur at around -0.7 V where the reduction starts to take place. The change in the gradient was found to be less prominent with graphene compared to that for the TiN electrode. On the return sweep, the reduction current was lower with two inflection points at -1.2 V and -0.7 V. The latter is presumably related to the cessation of $WS_2$ deposition.

The $WS_2$ deposition on graphene micro-electrodes was performed by fixing the potential at -1.7 V and changing the deposition time to achieve the desired thickness. The first deposition was performed for a 5 min period. Following the deposition, the substrate was rinsed in pure $CH_2Cl_2$. Figure 2 (c) shows an image of the deposited $WS_2$ films over the graphene electrodes depicted by bright patterns of zigzag and straight lines as per the mask design. In comparison, a graphene film that has not been connected to the gold pad appear in its original grey colour as shown in Figure 1, confirming there is no $WS_2$ deposition on these regions of the graphene film nor the presence of residues from the electrolyte. The as-deposited $WS_2$ film is amorphous, as shown by the absence of a Raman signal from these films. Therefore, an annealing step was performed to crystallize the film. The electrodeposited film was annealed in a tube furnace at 500 ºC for 2 h at 0.1 mbar. The sample was annealed in the presence of sulfur powder in the tube furnace to compensate for the loss of (volatile) sulfur from the film during the process. Thermal annealing was found to significantly improve the crystallinity of the material as shown in the Raman results below.

## Material Characterisation

### Atomic Force Microscopy

Atomic force microscopy was used to characterize the thickness of the $WS_2$/graphene heterostructure and study the uniformity and continuity of the deposited films. Figure 2 (d) and (e) show images of two different films grown for 5 and 1 min, respectively. The 5 min deposition shows a total step height thickness of 27.3 nm, while the 1 min deposit shows a step height thickness of 9.5 nm, see Figure 2 (f). These values correspond to the sum of the step height created in $SiO_2$ due to over etching, the thickness of the graphene electrode, and the thickness of the deposited $WS_2$ film. Supporting Figure S3 shows more AFM data taken from the 5 min and 1 min films along patterned zigzag lines.

Aiming towards monolayer $WS_2$, we further reduced the deposition time to 10 s, which led to a step height of 4.4 nm, as illustrated in Figure S4 and Figure 2 (f). Since the step height before the deposition was measured to be 2.3 nm, the actual thicknesses of the deposited $WS_2$ for 5 min, 1 min and 10 s are 25 nm, 7.2 nm and 2.1 nm, respectively. The layer-to-layer

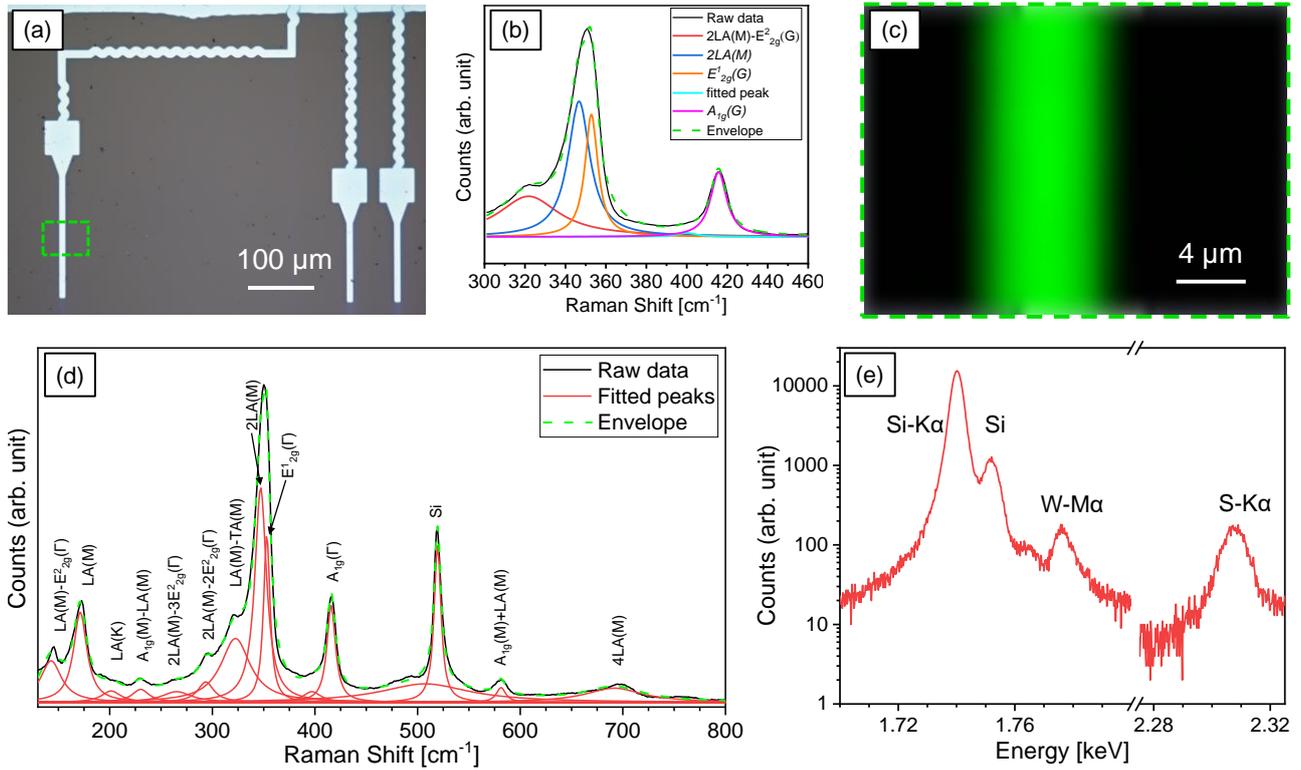

Figure 3 (a) An optical microscope image of patterned $WS_2$/graphene heterostructure. (b) A typical Raman spectrum taken from the $WS_2$ deposits showing the signature of the material and the fitted peaks associated with various vibrational modes in the material. (c) A Raman spatial map from the green squared region shown in (a) plotted as function of the intensity of the peak that is composed of the 2LA(M) and $E^1_{2g}$ peaks. (d) A wider Raman spectrum along with the fitted peaks showing the common vibrational modes for $WS_2$ and Si for the range 100-800 cm$^{-1}$. (e) A WDX spectrum of deposited $WS_2$ showing the Mα and Kα energy signatures of W and S, respectively.

spacing in $WS_2$ has been measured before to be approximately 0.65 nm.[4, 17] Hence the 10 s deposit is estimated to have produced a maximum of trilayer $WS_2$ thickness. The RMS roughness of the 10 s deposit is 0.28 nm. By plotting a graph of the measured thickness vs deposition time, the electrodeposition of $WS_2$ was found to behave linearly with a deposition rate of $0.10 \pm 0.03$ nm/s, see supporting Figure S5. The AFM images of the heterostructures show that the electrodeposited films have very good uniformity and continuity across the patterns. The uniformity remains high even for ultra-thin films, including the 10 s deposits. Compared to previous reports of electrodeposited $WS_2$ that mostly show very rough and discontinuous films, this work represents an important advancement in $WS_2$ production via electrodeposition, making it a competitive alternative deposition technique.[28–30] It is also worth noting that each of these depositions was performed using solutions prepared on different days, using different batches of synthesised precursor, indicating good repeatability for this process.

**Raman and Energy Dispersive X-ray Spectroscopy**

Raman spectroscopy is a common technique for characterizing the physical properties of 2D materials and TMDCs. In this work, the presence of $WS_2$ on the substrate, its uniformity and degree of crystallinity were investigated by measuring the Raman scattering of a 532 nm wavelength laser at room temperature. Light excitation and collection was done via a 50x objective lens which can reduce the laser spot diameter to 1 μm to probe the patterned areas. The consequence of using an excitation wavelength of 532 nm is that the Raman spectrum becomes enriched with second order harmonic modes.[42] Figure 3 (a) shows an optical microscope image of patterned $WS_2$/graphene heterostructures and a dashed line region where Raman spectra and maps were recorded. Figure 3 (b) presents a typical spectrum obtained from electrodeposited $WS_2$ over graphene microelectrodes. Variations in the intensity were observed depending on the thickness of the deposited material. Multi-peak Lorentzian fitting was applied to deconvolute the various peaks from the spectrum. The Raman shift of the $E^1_{2g}$ (Γ) peak, which relates to the in-plane vibration of W and S atoms is at 352.9 cm$^{-1}$, while the Raman shift of the $A_{1g}$ peak which involves the out-of-plane displacement of S atoms is 415.8 cm$^{-1}$. The shift difference between the $E^1_{2g}$ and the $A_{1g}$ peaks is 62.9 cm$^{-1}$ and the intensity ratio between the 2LA(M) and the $A_{1g}$ peaks (2LA(M)/$A_{1g}$) was calculated to be 1.94. A Raman map was recorded by scanning the region at 1 μm steps as shown in Figure 3 (c). The map is plotted by taking the spatial intensity of the raw signal peak at 350 cm$^{-1}$. It is clear to see that the map intensity is uniform across the patterned film which indicates that the quality of the film is highly uniform. A wider Raman spectrum of the $WS_2$ deposit is presented in Figure 3 (d) showing the presence of all the previously reported peaks from $WS_2$ with sharpness that is similar to $WS_2$ grown by CVD or via $WO_3$ sulfurization, indicating the high quality of the material.[42, 43]

A common technique to quantify the compositon of a material is to use energy dispersive X-ray spectroscopy. However, due to the close spectral energy between the Si- Kα and W-Mα electron energy peaks wavelength dispersive X-ray (WDX) spectroscopy was employed instead. Figure 3 (e) shows a WDX spectrum obtained from a film deposited for 5 min which had a thickness of roughly 27 nm. The electron energy peaks of W Mα and S Kα peaks were found at 1.776 eV and 2.307 eV, respectively. The S/W composition ratio was quantified to $2.0 \pm 0.1$.

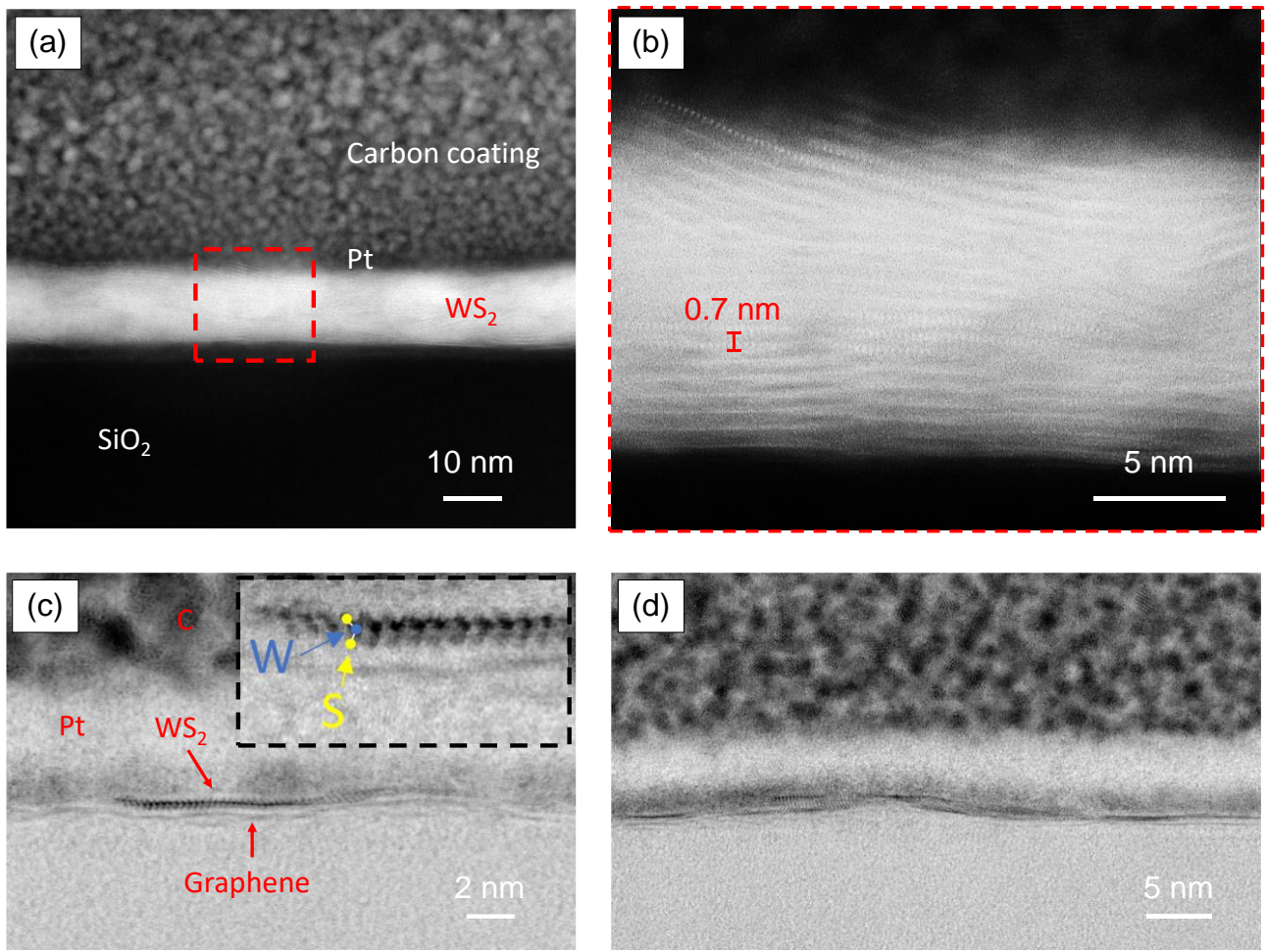

Figure 4. (a) Cross-sectional annular dark-field TEM image of WS$_2$ electrodeposited for 5 min on patterned graphene electrodes. (b) High magnification image of the highlighted region in (a) showing the layer ordering of the 2D WS$_2$ growing preferentially in the horizontal direction along the surface of the substrate. (c) and (d) Bright field TEM images of a film grown for 10 s showing regions of electrodeposited films composed of 1-2 layers over graphene.

**Transmission Electron Microscopy**

To provide further insight on the physical nature of the film, TEM imaging was performed by taking a lamella slice of the patterned film using a focused ion beam miller. Pt and C were deposited over the film to protect it during the milling process. Figure 4 (a) presents an image taken via annular dark field (ADF) mode showing the WS$_2$ film in between the SiO$_2$/Si substrate and the protection layers. A higher magnification image of the highlighted region in (a) is presented in Figure 4 (b). The image clearly shows the preferential crystallization of the WS$_2$ layers in the horizontal direction above the substrate. The vertical layer-to-layer distance in the stack was measured to be 0.7 ± 0.1 nm which matches very well with previous work.[42] Overall, it is clear that the film is composed of well-ordered 2D layers and the layer ordering of a particular layer tends to follow that of previous layers. It was noticed that the horizontal layer ordering changes slightly in some regions where the thickness is higher.[44] Figure 4 (c) and (d) show high resolution TEM images taken via bright field for a film grown for 10 s. These films are observed to be composed of 1-2 layers only. It is clear to note here the presence of the graphene layer under the WS$_2$ film which matches TEM images of graphene layers from the literature.[45] The grown WS$_2$ was also noted to clearly follow the topography of the graphene film. The inset of Figure 4 (c) show a digital magnification of the atomic structure of the WS$_2$ film. With a careful look at the inset, the reader should be able to distinguish the W and S atoms in the lattice. We have overlayed a schematic of one molecule to make our indication clearer. The W atoms appears darker in BF compared the S atoms due to their difference in atomic weights.

**X-ray Photoelectron Spectroscopy**

X-ray photoelectron spectroscopy (XPS) measurements were performed on the electrodeposited WS$_2$ films to study the effect of annealing the film and to quantify its W:S composition. Figure 5 shows the XPS scans taken at the binding energy ranges of W and S on the as-deposited film (a and b) and after annealing (c and d). A typical W 4 f spectrum from WS$_2$ is composed primarily of a spin doublet corresponding to 4 f$_{5/2}$ and 4 f$_{7/2}$.[46] Here, these were observed at 32.9 eV and 34.9 eV, respectively. The energy of these peaks corresponds to the 2H phase of WS$_2$. Another prominent spin doublet (red) was also found at higher energies. The second doublet observed here has been reported by previous studies to be attributed to oxidation, resulting in W$^{6+}$ peaks.[47, 48] These could be related to surface oxidation of the film resulting in the formation of WO$_3$.[17, 49, 50] After annealing, the intensity of the W$^{6+}$ peaks reduce significantly and an enhancement in the W$^{4+}$ peaks was observed with a significant reduction in their full width half maximum (FWHM). This clearly indicates an enhancement in the crystallinity of the film in the 2H phase. The evolution of the sulfur S 2p (blue) peaks was similar to that of the W$^{+4}$ 4f peaks after annealing. The S 2p$_{1/2}$ and 2p$_{3/2}$ peaks were measured at 161.9 eV and 163.1 eV, respectively. A clear reduction in the FWHM was also found here, with an increase in the overall intensity of the peaks. A second doublet was also observed (red) which was attributed to $S_2^{2-}$ terminal sulfur bond formation that might have resulted from extra sulfur deposited on the film's surface during the annealing process.[51]

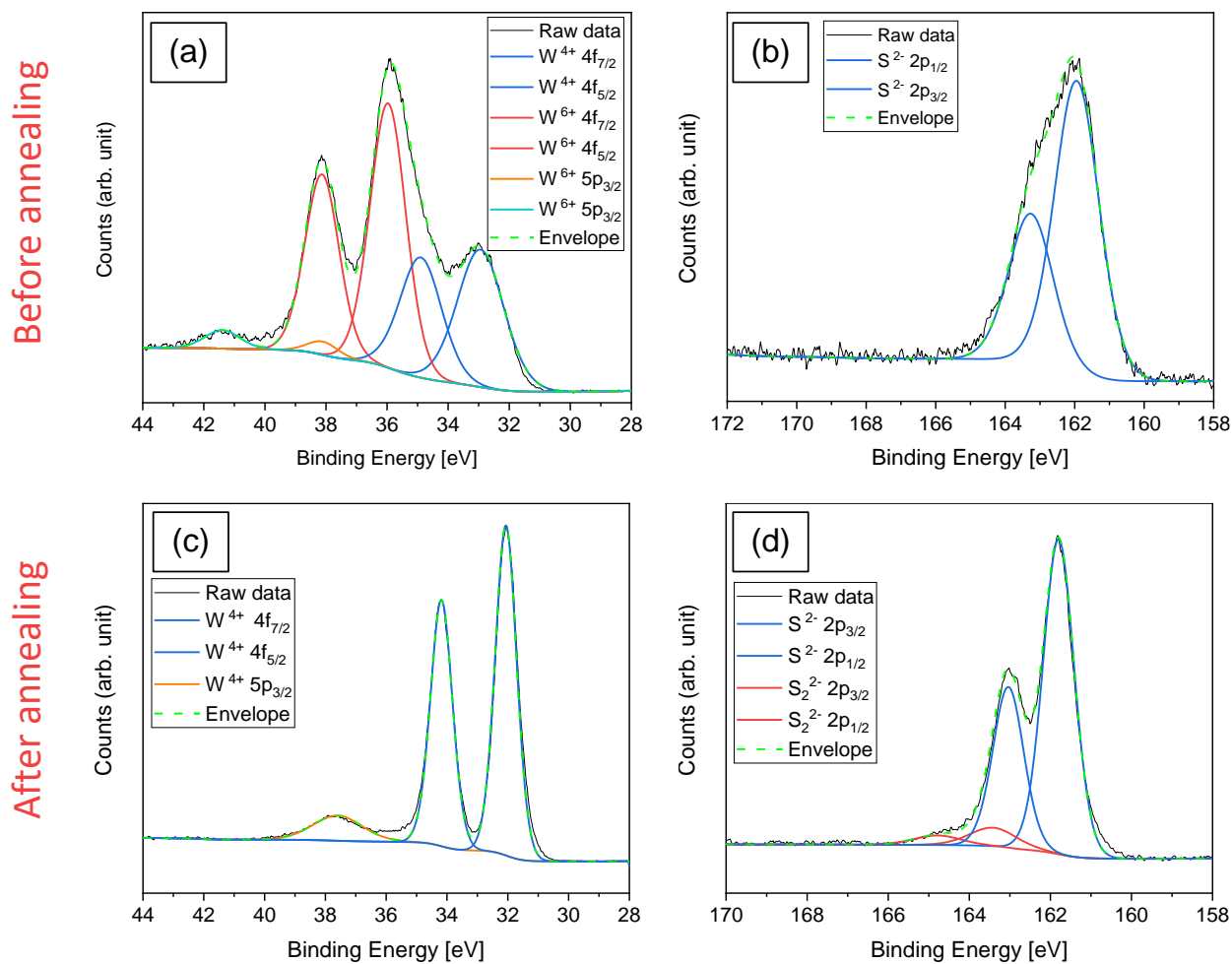

Figure 5 XPS measurements across the W (a) and S (b) binding energy range for the film before and after (c and d) annealing. The graphs show clear enhancement in the material crystallinity after annealing as evidenced by the reduction in the FWHM of the emission peaks and the increase in the intensity.

## Conclusions

The electrodeposition of $WS_2$ over graphene electrodes was demonstrated using a single source precursor. By fabricating graphene micro-electrodes, we were able to demonstrate selective growth of $WS_2$ films into predefined arbitrary micro-patterns. The electrodeposition of $WS_2$ was found to be controllable and follows a linear behaviour. AFM and Raman spectroscopy confirmed that the electrodeposited materials are highly uniform and of good crystallinity after annealing. High resolution TEM imaging has shown that the films are composed of well-ordered stacks of 2D layers of crystalline $WS_2$. TEM imaging also confirmed that this method is capable of producing films that are as thin as 1-3 layers. XPS measurements showed the influence of annealing the film on its stoichiometry and crystallinity, aligning with previous studies of $WS_2$ films grown using other techniques. This work demonstrate the potential of electroplating to enable 2D heterostructures at the microscale using graphene patterned electrodes.

## Author contributions

YJN fabricated the electrodes and performed Raman and XPS characterisation; ST performed electrodeposition experiments; SR grew and transferred the graphene electrodes prior to fabrication; VG prepared the precursors and annealed the films, NA and JZ performed AFM microscopy, YH performed TEM microscopy. CHDG, PNB, GR, NK, RB and ALH supervised the project. YJN wrote the manuscript with contributions from all authors.

## Acknowledgements

The research work done in this article was financially supported by the Engineering and Physical Sciences Research Council (EPSRC) through the research grant EP/P025137/1 (2D layered transition metal dichalcogenide semiconductors via non-aqueous electrodeposition) and the programme grant EP/N035437/1 (ADEPT - Advanced devices by electroplating).